\documentclass[12pt]{article}
\usepackage{tensor, amsmath}

\usepackage{hyperref}
\hypersetup{
    colorlinks,
    citecolor=red,
    filecolor=black,
    linkcolor=blue,
    urlcolor=black
}

\begin{document}

\title{\textbf{Clarifying possible misconceptions in the foundations of general relativity}}

{
\renewcommand{\thefootnote}{\fnsymbol{footnote}}

\date{}
\author{Harvey R.~Brown\footnote{harvey.brown@philosophy.ox.ac.uk}~ 
and
James Read\footnote{james.read@philosophy.ox.ac.uk}
}

\maketitle
\begin{center}
\emph{Faculty of Philosophy, University of Oxford,}\\
\emph{Radcliffe Humanities, Woodstock Road, Oxford OX2 6GG, UK.}

\par\end{center}

}

\begin{abstract}
We discuss what we take to be three possible misconceptions in the foundations of general relativity, relating to:~(a) the interpretation of the \emph{weak equivalence principle} and the relationship between gravity and inertia; (b) the connection between gravitational redshift results and spacetime curvature; and (c) the \emph{Einstein equivalence principle} and the ability to ``transform away" gravity in local inertial coordinate systems. 
\end{abstract}

\tableofcontents{}

\section{Introduction}

Einstein's theory of general relativity (GR) is often claimed to be the most striking, profound, and beautiful physical theory ever constructed. It is undoubtedly conceptually rich;~one hundred years after its development, GR continues to inspire not only spectacular tests of its predictions,\footnote{For the recent detection of gravitational waves, see \cite{Abbott}.} but also debate about its foundations. 
In this paper, we describe and resolve what we take to be three possible misconceptions in the foundations of GR, which in our experience may hinder students' grasp of the fundamental concepts of the theory. 

The first of these possible misconceptions pertains to the so-called \emph{weak equivalence principle} (WEP) in the version which claims that one can ``simulate'' a (homogeneous) gravitational field by way of a uniform acceleration. Our gripe is not the common one that real gravitational fields are not strictly homogeneous. Rather, the putative misconception consists in viewing the WEP as expressing the \emph{equivalence} of (approximately homogeneous) gravity and inertia. In fact, in GR neither of these terms is fundamental, and if anything the theory implies the \emph{reduction} of this kind of gravity to inertia. In section \eqref{s1}, we provide a discussion of the merits of this way of thinking about the WEP.

The second possible misconception relates to the notion that gravitational redshift experiments provide evidence for spacetime curvature. They do, but contrary to what is claimed in a couple of important modern textbooks on GR, a single gravitational redshift experiment \emph{does not} require an explanation in terms of curvature. Rather, it is only \emph{multiple} such experiments, performed at appropriately \emph{different} locations in spacetime, that suggest curvature, via the notion that inertial frames are only defined locally. In the process of elaborating on this in section \eqref{s2}, we also take a swipe at the nomenclature associated with the ``clock hypothesis''.

The third possible misconception has to do with the idea, still occasionally rehearsed in the  modern literature, that relative to local Lorentz frames, the dynamical equations of GR associated with the non-gravitational interactions do not contain curvature-related terms; that for matter fields, gravity can be locally ``transformed away''. In fact, this is \emph{not} generally the case, making the local validity of special relativity in GR -- sometimes called the \emph{strong} or \emph{Einstein equivalence principle} -- a subtle matter. Textbooks sometimes contain both the ``transforming away'' claim and the details (related to second order equations) that refute it! Such issues are discussed by way of explicit example in section \eqref{s3}.

Though all three of the above points rarely inhibit the successful performance of calculations in GR, they go to the heart of our understanding of notions crucial to the theory, such as gravitation, inertia, spacetime curvature, and the connection between GR and the special theory.
We thus hope that this paper will have pedagogical merit for students of this most remarkable and subtle of theories, as our understanding of ``gravity'' continues to grow and evolve in the years beyond the centenary of GR.

\section{The Weak Equivalence Principle}\label{s1}

\subsection{Einstein}\label{Einstein}

In 1907, when thinking about someone unfortunate enough to be falling off a roof, Einstein had the happiest thought of his life (``gl\"{u}cklichste Gedanke meines Lebens'').\footnote{\cite{E2}.} He realised that in the immediate vicinity of such an observer, gravity would seem to disappear. This, in itself, was not a revolutionary insight; it might occur to anyone familiar with Newtonian gravity.\footnote{Newton's own appreciation of this fact is seen in his application of Corollary VI of the laws of motion to gravitation in the \textit{Principia}, particularly in relation to the discussion of the system of Jupiter and its moons. For a recent discussion, see \cite{Saunders}.} But Einstein recalled the important role that recognition of the coordinate-dependent nature of electric and magnetic field strengths played in his development of special relativity, and concluded that the gravitational field likewise ``has only a relative existence''. This would be the key to incorporating gravity into the relativistic framework, or some modification thereof.

The curious nature of Einstein's implementation of this idea has arguably given rise to a long-standing confusion about gravity in the literature. He noted that inertial effects detected by observers accelerating relative to (Newtonian) inertial frames are likewise frame-dependent, and it occurred to him that somehow inertia and gravity are ``identical''. Consider the centrifugal force associated with rotation:
\begin{quotation}
... the same property which is regarded as \textit{inertia} from the point of view of a [inertial] system not taking part in the rotation can be interpreted as \textit{gravitation} when regarded with respect to a system that shares the rotation.\footnote{\cite{E21}.}
\end{quotation}
The problem, as Einstein saw it, was that in Newtonian mechanics inertial ``fields'' are not ``real''; a replacement of Newton's gravitational theory might resolve the matter. 

The replacement came with the field equations of 1915, and in a 1916 publication Einstein again stressed the relative nature of gravity and inertia. He suggested that gravity, in particular, should be represented by the term in the geodesic equation for the motion of test particles that depends on the components of the affine (Christoffel) connection,\footnote{\cite{E1}.} thus highlighting the coordinate-dependent nature of gravity. Einstein was still committed to the idea of unifying gravity and inertia in a manner reminiscent of the way Maxwell theory unifies electricity and magnetism\footnote{For a detailed discussion of Einstein's interpretation of GR, see \cite{Janssen} and \cite{Dennis}. It is noteworthy that the covariant form of Maxwell's equations also played an important role as a template in the technical development of the field equations themselves (see \cite{Janssen-Renn}).} -- and unification in this context does not mean equivalence, despite Einstein's use of the word ``identical''.  His 1916 suggestion is rather unusual by today's lights,\footnote{As Peter Havas noted in 1967, starting with coordinates in which the connection coefficients vanish at a spacetime point, the mere transformation to non-Cartesian coordinates may make the term in the geodesic equation containing the connection reappear at that point (see \cite[p.~134]{Havas2}). And according to Einstein's suggestion, this makes gravity appear too, even though no acceleration is involved relative to the initial local inertial frame.}  but the thorny issue of the relationship between gravity and inertia that Einstein raised lives on in the literature. Before we proceed to discuss this, it is worth highlighting another important remark Einstein made in the same 1916 publication, namely that ``labels'' like gravity and inertia are ``in principle unnecessary'' in GR, but that ``for the time being they do not seem worthless to me, in order to ensure the continuity of thoughts''.\footnote{\cite{E1}.} He would make essentially the same point in 1922.\footnote{\cite{E22}.}

\subsection{Equivalence?}\label{equivalence}

Such continuity of thoughts in going from Newtonian gravity to GR may have its merits, but it can lead to confusion, and arguably nowhere is this more apparent than in relation to what has come to be called in the literature the \emph{weak equivalence principle} (WEP). Recall that in Newtonian gravity, for Galileo's law of universal free fall to hold (namely that in the absence of air resistance, all bodies fall to the Earth at the same rate, independent of their constitution), inertial mass and gravitational mass must be proportional.
It is widely considered that this ÒequivalenceÓ between quite distinct concepts is \textit{a priori} surprising; indeed removing the mystery was one of the motivations behind Einstein's search for a new theory of gravity.
In GR, however, gravitational and inertial mass are no longer fundamental notions, and accordingly the WEP is sometimes couched directly, and more generally, in terms of the motion of test bodies. Let us start with the version found in the recent textbook by Erik Poisson and Clifford Will:
\begin{description}
\item[WEP 1:] If a test body is placed at an initial event in spacetime and given an initial velocity there, and if the body subsequently moves freely, then its world line will be independent of its mass, internal structure and composition.\footnote{\cite[p.~218]{Will}. For a review of the experimental evidence in favour of WEP 1, see  \cite[ch.~3]{Ciuf} and \cite{Giulini}; in the latter the principle is called The Universality of Free Fall (UFF).} 
\end{description}
Within GR, this principle is incontestable, ``test'' body being understood in the usual way and hence excluding spinning particles.  Worldlines of test bodies are identified in the theory with time-like geodesics of the metric,\footnote{This identification is sometimes called \textit{Einstein's strong equivalence principle} (see e.g.~\cite[p.~601]{TW}), not to be confused with the similarly-named principle discussed in section \eqref{s3} below. Initially, Einstein introduced such geodesic motion as a postulate, but by 1927 he and Grommer realised what others like Eddington had already seen, that it can be regarded as a consequence of the Einstein field equations of GR. An important qualification to this claim will be indicated in section \eqref{s3} below. (For a history of the geodesic theorem for test bodies, see \cite{Havas}.) Note that the extent to which the geodesic principle can be derived from first principles in GR depends to some extent on the chosen formulation of GR. In the case of the trace-free version of the field equations, the vanishing of the covariant divergence of the stress-energy tensor, a crucial element in the standard proof of the geodesic theorem, is itself a postulate and no longer a consequence of the field equations involving the metric (see \cite{Ellis}).} which of course are independent of the constitution of the bodies. This explains why  for a freely falling observer, the motion of other nearby falling objects would appear to be inertial and, hence, ``independent of their special chemical or physical nature'', in Einstein's words.\footnote{\cite{E2}.} And for an observer accelerating relative to freely falling frames, such as Galileo on the surface of the Earth, freely falling objects would appear to accelerate independently of their constitution. But now there are no forces (in the sense of Newton) acting on the falling bodies!

For the present, the important thing to note about WEP 1 is that it makes no mention of the ``gravitational field''. This is not true for an alternative version of the WEP, and expressed in the 2003 textbook due to Hartle:\footnote{\cite[p.~113]{Hartle}.}
\begin{description}
\item[WEP 2:] Einstein's equivalence principle is that idea that \textit{there is no experiment that can distinguish a uniform acceleration from a uniform gravitational field.} The two are fully ``equivalent''. (original italics)\footnote{Though WEP 1 and WEP 2 are both referred to as the ``weak'' equivalence principle, we stress that we do \emph{not} conflate the two. In fact, in what follows we will argue broadly in favour of WEP 1, and against WEP 2, from the perspective of GR.}
\end{description}
Whether this idea does justice to Einstein's original thinking is doubtful, for the reasons given above. But statements similar to WEP 2 (albeit not always labelled as such) can be found in many prominent texts;\footnote{See, e.g., \cite[p.~14]{Gron}, \cite[p.~68]{Weinberg}, \cite[p.~271]{Zee} and \cite[p. 49]{Carroll}. Such language can also be found in foundationally-oriented papers, such as \cite{Nobili}.} others -- in line with WEP 1 -- avoid any such formulation.\footnote{See, e.g., \cite{MTW} and \cite{Wald}.}

The key issue here is what \textit{equivalence} means in  WEP 2  (note the use of scare quotes by Hartle). One use of  WEP 2 is in predicting, say, the behaviour of light in the Earth's gravitational field, in a pre-GR theoretical scenario which is either Newtonian (as Hartle happens to be considering\footnote{\textit{op.~cit.}}) or special relativistic. In the first case, there are equations for gravity but not for light; in the second case it is the other way round. In either case, WEP 2 is being employed as a heuristic principle, or \emph{ansatz}, within a theory of limited means, possibly to pave the way for GR. The question we are interested in is what to make of WEP 2 once GR is adopted. It is the change of the status of the principle in the transition to GR that is frequently overlooked in the literature. Within GR, 
WEP 2 is liable to be misleading, as it might suggest that uniform acceleration and homogeneous gravitation in GR are {distinct notions}, rather in the way inertial and gravitational mass are in Newtonian gravity. Indeed, this leads us to identify what we take to be the \emph{first possible misconception of GR}:

\begin{description}
\item[Misconception 1:] Uniform acceleration and homogeneous gravitational fields in GR are {conceptually distinct}, but somehow empirically equivalent.
\end{description}
Here, there is a risk of inferring that uniform accelerations, \textit{in the absence of a gravitational field}, can, for some reason, ``simulate'' a \textit{real} gravitational field. Thinking in this manner can incline the student of GR to ask questions such as:~``{Where} is the gravitational field? How do I separate the components of a particle's acceleration associated with gravitation and with the uniform acceleration of the reference frame?''

The confusion arises because the (locally) homogeneous gravitational field being referred to is a Newtonian notion which only really makes sense when the inertial frames themselves are not freely falling. But a fundamental principle of GR is that  inertial frames are redefined so as to reduce the non-tidal component of what Newton called free fall under gravity to inertia (or rather geodesic motion), as observed from accelerating observers. In other words, in GR, \emph{inertial frames are freely falling frames}, thereby in an important sense \emph{reducing} gravity to inertia, and certainly removing its status as a force. (This point will be made clearer in the discussion of the strong equivalence principle in section \eqref{s3} below.)
Rather than ``equivalence", it would be more helpful to use the term ``reduction" in the context of WEP 2 insofar as it relates gravity to inertia. Perhaps better still would be to avoid using the terms ``gravity'' and ``inertia'' altogether -- as Einstein hinted in 1916, and again in 1922. The frequent appearance of ``gravitational field'' in treatments of GR, unless it is associated with spacetime curvature, can be just as misleading as the use of the classical term ``particle'' for denoting an electron, for instance, in quantum mechanics.

Indeed, a common remedy is precisely to restrict the term gravity to refer to inhomogeneous fields, i.e.~non-vanishing spacetime curvature. Such is the approach taken recently by Poisson and Will, who remind us that in the real world, strictly homogeneous (uniform), static gravitational fields don't exist. Thus, in discussing what they take to be Einstein's principle of equivalence (essentially WEP 2), they write:
\begin{quotation}
... his [Einstein's] formulation is deeply flawed when taken literally. ... the strict adoption of his principle has led to a pointless literature of apparent paradoxes, debates and conundra. ... a uniform gravitational field is not a gravitational field at all. It is the ``field'' experienced by an observer undergoing constant acceleration in Minkowski spacetime, and nothing else. ... Einstein's original formulation is not an equivalence but a tautology.\footnote{\cite[p.~221]{Will}. Rohrlich (\cite{Rohrlich}, p. 186) likewise considered WEP 2 to be without empirical content, but he saw it as a definition of inertial frames.}
\end{quotation}
But whatever Einstein meant, his reasoning was not intended to be restricted to Minkowski spacetime; it was intended to provide a way out of it. Indeed, the assumption that WEP 2 holds in general, and not just special, relativity is ubiquitous in text books. Our point is that reasoning similar to that of Poisson and Will applies to an \textit{apparently} uniform gravitational field as seen by an observer doing local experiments (i.e. ones insensitive to tidal effects) in curved spacetime. We intend to make this notion of locality more precise in the following sections. In the meantime, 
we suggest a correction to the first possible misconception as follows:

\begin{description}
\item[Correction 1:] Inertia and gravity are not fundamental notions in GR, and in so far as the terms can be used, (approximately homogeneous) gravitational fields are reduced to, and not simulated by, inertial effects.  
\end{description}

\section{Gravitational Redshift and Spacetime Curvature}\label{s2}

\subsection{Redshift}

What is the experimental evidence for the  non-Newtonian claim made in GR that inertial frames are freely falling? As Kip Thorne and Clifford Will stressed in 1971, such evidence does not come from elementary particle physics, which allows for accurate testing of the validity of the Lorentz group, but from a series of experiments starting in the 1960s, associated with ``gravitational redshift''.\footnote{See \cite[p.~602]{TW}, and \cite[pp.~1055ff.]{MTW}. In the former, it is argued that despite not relying on the details of the Einstein field equations, the redshift experiment should not be regarded as a ``weak test'' of GR.}  The most celebrated of these -- for better or for worse -- is the experiment performed by Robert Pound and Glen Rebka at Harvard University in 1960.\footnote{A useful historical account of the early redshift experiments is found in \cite{Hentschel}, which casts some doubt on whether the earlier experiment performed by J.~P.~Schiffer and his collaborators at Harwell in England was as inconclusive as Pound claimed at the time. For more recent and more accurate versions of the redshift experiment, see the review in \cite[p.~103]{Ciuf}.} Such high precision experiments were made possible by the discovery in 1958 of the M\"{o}ssbauer effect, which allowed for $\gamma$-rays in a certain narrow frequency range to be emitted and absorbed by solid samples containing radioactive Fe$^{57}$. When two such samples are placed vertically with a height difference $h$ (and so at a difference $gh$ in the gravitational potential in the language of Newton), GR predicts that photons emitted from one sample will no longer be absorbed by the other.  But if the absorber is put into a certain degree of vertical motion relative to the source, the resulting Doppler effect can restore absorption.

This ``redshift'' effect follows directly from the claim that the emitter and absorber are accelerating vertically at a rate of $g$ ms$^{-2}$ relative to the (freely falling) inertial frames. Indeed a simple calculation, variants of which are standard fare in GR textbooks\footnote{See e.g.~\cite[pp.~133ff.]{Hartle}.} (so the details are not repeated here), shows that\footnote{As noted in \cite{Landsberg1, Landsberg2}, in some textbooks this result is erroneously derived from models in which the detector moves \emph{with respect to} the source (see e.g.~\cite[p.~190]{MTW}). In fact, this is not an example of a gravitational redshift-type result, but rather of the relativistic Doppler effect. In a correct derivation of \eqref{gredshift}, source and accelerator must be considered at rest with respect to one another, but accelerating with respect to an inertial frame.}
\begin{equation}\label{gredshift}
\Delta t_{B}\simeq\left(1- \frac{gh}{c^{2}}\right)\Delta t_{A}.
\end{equation}
where $\Delta t_{A}$ is the coordinate interval between successive electromagnetic wave crests being emitted by sample $A$, and $\Delta t_{B}$ is similarly defined for reception at  the lower sample $B$.\footnote{Note that if the source $A$ is the higher sample, then received signals at $B$ will in fact be \emph{blueshifted}. If the positions of emitter and receiver are swapped, then the situation resembles the original Pound-Rebka experiment \cite{PR2}, and the received signals (at the higher sample) will indeed be \emph{red}shifted (see e.g.~\cite[pp.~187, 1056]{MTW}). For details concerning later more accurate versions of the redshift experiment, see \cite[ch.~3]{Ciuf}.} Note that these quantities are coordinate intervals defined relative to an inertial frame, and according to special relativity cannot be identified in general with proper times associated with accelerating ``clocks''. But the calculation relies on a number of approximations that, amongst other things, render negligible special relativistic effects such as time dilation. 
Two further assumptions are worth emphasising:~that the speed of light relative to the inertial frame in question is independent of the speed of the emitter,\footnote{This is of course essentially a special case of Einstein's 1905 light postulate \cite{E3} (for discussion, see \cite[ch.~5]{Brown}). But note that the inertial frame is now freely falling!} and that the Fe$^{57}$ nuclei used in practical experiments such as the Pound-Rebka setup satisfy the \textit{clock hypothesis} for accelerating clocks (see section \eqref{clock} below).

Now consider the rest frame of the surface of the Earth, i.e.~comoving with the laboratory. Since the ``gravitational field'' seen in this frame is static, the relevant coordinate intervals at $A$ and $B$ must be equal:~$\Delta t'_A = \Delta t'_B$,\footnote{See, e.g.~\cite[p.~602]{TW}.} so insofar as $\Delta t_A$ and $\Delta t_B$ above are approximately proper times, it appears that clocks run at different rates depending on their height, or the gravitational potential $\Phi$. Indeed, it is common to see equation \eqref{gredshift} rewritten as 
\begin{equation}
\Delta t_{B}\simeq\left(1- \frac{\Phi_{A}-\Phi_{B}}{c^{2}}\right)\Delta t_{A}
\end{equation}
given that  $\Phi_{A}-\Phi_{B}= gh$; due recognition should be given, though, to the dangers associated with a literal reading of the Newtonian gravitational potential.

Note that this derivation of the (first order) redshift does not depend on the nature of the clocks at $A$ and $B$, as long as they satisfy the clock hypothesis under acceleration. Redshift is a \textit{universal effect} in this restricted sense. Furthermore, the derivation makes no appeal to the Einstein field equations in GR.\footnote{See \cite{Schild1}.} In fact, there is nothing incomplete about the derivation, and the temptation to provide a deeper explanation of the redshift phenomenon has sometimes led to confusion, as we see now.  
 
\subsection{Curvature}\label{3.2}

In his 2003 textbook, James Hartle argues that to explain the universal redshift on the basis of some kind of non-geometrical ``gravitational'' effect is akin to committing the following fallacy. Imagine supposing the surface of the Earth is flat and explaining its apparent curvature by appealing to a universal effect on rulers, namely that they universally somehow become longer the further they are from the equator. In Hartle's words:
\begin{quote}
It is simpler,
more economical, and ultimately more powerful to recognise that distances
on Earth are correctly measured by rulers and that its surface is
curved. In the same way, it is simpler, more
economical, and ultimately more powerful to accept that clocks correctly measure timelike distances in spacetime and that its geometry is curved.\footnote{\cite[p.~126]{Hartle}.}
\end{quote}

Sean Carroll, in his  2004 textbook, arrives at a similar conclusion regarding redshift:
\begin{quotation}
... simple geometry seems to imply that the [emission and reception intervals] must be the same. But of course they are not; the gravitational redshift implies that the elevated experimenters observe fewer wavelengths per second ... We can interpret this roughly as `the clock on the tower appears to run more quickly'. What went wrong? Simple geometry -- the spacetime through which the photons traveled was curved.

We therefore would like to describe spacetime as a kind of mathematical structure that looks locally like Minkowski space, but may possess nontrivial curvature over extended regions.\footnote{\cite[pp.~53-54]{Carroll}. For similar conclusions, see e.g.~\cite[p.~210]{Ohanian}, \cite[p.~27]{Rindler} and \cite[p.~189]{MTW}.}
\end{quotation}

But the derivation outlined above nowhere depends on tidal effects; so although, as we shall see, there is a connection between curvature and redshift experiments, it is not this one. In particular, what Carroll means by an ``extended region'' of spacetime is not exemplified by the Pound-Rebka experiment.
Indeed, with these assessments of gravitational redshift results in mind, we can now state what we take to be the \emph{second possible misconception of GR} as follows:

\begin{description}
\item[Misconception 2:] An explanation for the results of a single gravitational redshift experiment of Pound-Rebka type will appeal to a notion of spacetime curvature.
\end{description}

To see clearly why this is indeed a misconception, consider the geometric version of the above derivation, in flat Minkowski spacetime with metric $\eta_{\mu\nu}$, again with the laboratory accelerating relative to the inertial frames by the arbitrary amount $a$ ms$^{-2}$.
Let $d \tau_A$ denote the \textit{exact} proper time associated with the emission of successive wave peaks at $A$, and similarly $d \tau_{B}$ the \textit{exact} proper time associated with the arrival of successive wave crests at $B$. Again, let the primed coordinates $x^{\prime\mu}$ be those associated with the frame comoving with the laboratory.\footnote{In what follows, we use the Einstein summation convention, with Greek
indices ranging from 0 to 3.}   So
\begin{eqnarray}
d\tau_{A}^{2} & = & \eta_{\mu\nu}\left(A\right)dx'^{\mu}_{A}dx'^{\nu}_{B} \label{eq3}\\
 & = & \eta_{00}\left(A\right)\left(dx'^{0}_{A}\right)^{2},
\end{eqnarray}
\noindent and similarly
\begin{equation}
d\tau_{B}^{2}=\eta_{00}\left(B\right)\left(dx'^{0}_{B}\right)^{2}.
\end{equation}
As noted earlier, for reasons of homogeneity of time, the coordinate intervals are equal: $dx^{\prime 0}_{A}=dx'^{0}_{B}$. Thus 
\begin{equation}\label{v2}
\frac{d\tau_{A}}{d\tau_{B}}=\sqrt{\frac{\eta_{00}\left(A\right)}{\eta_{00}\left(B\right)}}.
\end{equation}
Now consider the analogue of equation \eqref{eq3} for inertial (freely falling) coordinates
\begin{equation}
d\tau_{A}^{2} =  \eta_{\mu\nu}\left(A\right)dx^{\mu}_{A}dx^{\nu}_{A} = c^2 dt^2_{A} - dx^2_{A} -dy^2_{A} - dz^2_{A},
\end{equation}
and similarly for $d\tau_{B}^{2}$, where emitter and absorber lie along the $z$-axis. Taking into account the form of the transformations between the inertial and accelerating coordinates in special relativity,\footnote{See \cite[pp. 602, 603]{TW}, \cite[p.~166]{MTW}, \cite[p.~132]{Hartle}, and \cite[p.~285]{Will}.} along with $dt'_A = dt'_B$, we get, when $B$ is at the origin of the accelerating coordinate system,
\begin{equation}\label{v2}
\frac{d\tau_{A}}{d\tau_{B}}=\sqrt{\frac{\eta_{00}\left(A\right)}{\eta_{00}\left(B\right)}} = 1+\frac{ah}{c^{2}}.
\end{equation}
Putting $a = g$, this exact result is consistent with (1) given the approximations made in its derivation. The geometric derivation of the redshift underlines the fact that relative to a non-inertial frame, the Minkowski metric no longer takes its standard diagonal form, and in particular in an accelerating frame $\eta_{00}\left(A\right)$ is not equal to $\eta_{00}\left(B\right)$. 
\emph{Curvature is clearly not involved.}\footnote{We do not agree, however, with a long-standing notion that the Pound-Rebka effect can be understood by combining special relativity with WEP 2 (see \cite[\S IV]{Nobili}). Apart from our concerns about the significance of WEP 2 outlined in section \eqref{equivalence} above, we would also emphasise that in the 1905 version of special relativity, Einstein was explicit that the inertial frames were the same as Newton's:~they are not freely falling. In this case, no redshift is predicted; see \cite{TW}, p. 602.}~\footnote{The arguments leading to our equation \eqref{v2} above are essentially those given by Thorne and Will (\cite[pp.~602, 603]{TW}); the analysis there is given in the context of a curved spacetime geometry, however with the curvature playing no part in the analysis. Given Hartle's claim above that redshift is ultimately connected to spacetime curvature, it is striking that in Problem 6, p.~132 (\emph{op.~cit.}), Hartle uses the transformations to accelerating coordinates to conclude that accelerating clocks in special relativity are affected in accordance with our equation \eqref{v2}, but leaves how this is related to ``the equivalence principle idea'' as a question. Exercise 5.1 in the Poisson-Will textbook \cite[p.~285]{Will} almost exactly reproduces Hartle's Problem 6, even repeating the same question. This section of the present paper can be read as an attempt to answer this question.}

As alluded to above, however, there \emph{is} nonetheless a connection between gravitational redshift results and spacetime curvature.
The original redshift experiments were carried out at the Lyman Laboratory of Physics at Harvard University and at the Atomic Energy Research Establishment at Harwell in England. In each case, the result is explained by assuming that inertial frames are freely falling, as we have seen. But relative to the global freely falling frames at Harvard a freely falling object at Harwell is not moving inertially, and \textit{vice versa}. \textit{Inertial frames must be defined locally} in order to be compatible with such manifest ``geodesic deviation'' caused by the spherical shape of the Earth.\footnote{The importance, from the point of view of establishing spacetime curvature, of considering multiple redshift experiments spread out on the surface of the Earth was stressed by Schild in 1962 \cite[\S19]{Schild2}. But Schild defines the WEP in an unusually complicated fashion because the laboratories in his account are, confusingly, accelerating with respect to the Earth, not with respect to the free-fall frames.} In a similar vein, consider the first test, carried out by Brault in 1962, of the redshift of spectral lines emitted on the surface of the Sun and received at Earth.\footnote{\cite{Brault}. An interesting account of Brault's work and its reception is found in \cite{Hentschel}.} Brault's results were consistent with the assumption that inertial frames are unaccelerated in relation to freely falling test bodies at all points along the photon trajectory. As Thorne and Will state, this could not be the case ``if there were a single global Lorentz frame, extending throughout the solar system and at rest relative to its center of mass!''\footnote{\cite[p.~603]{TW}.} The fact that inertial frames can only be defined locally -- the essence of spacetime curvature -- is exposed not by a \emph{single} redshift experiment, but by \emph{multiple} redshift experiments sufficiently far apart. We can thus state a correction to the second misconception as follows:

\begin{description}
\item[Correction 2:] The results of a single gravitational redshift experiment of Pound-Rebka type do not require an explanation in terms of spacetime curvature. \emph{Multiple} gravitational redshift experiments performed at different spatial locations, however, require for their joint explanation the rejection of the global nature of inertial frames. Only in this case is one naturally led to the notion of spacetime curvature.
\end{description}
GR of course incorporates spacetime curvature, but again what is doing the work here in explaining an individual redshift effect is acceleration relative to the local freely-falling inertial frames. The precise nature of these frames, and the meaning of the claim that spacetime is locally Minkowskian, will be spelt out in
section \eqref{s3} below. We conclude this section by observing that if the position is taken that ``gravity'' is strictly spacetime curvature, individual redshift effects such as exemplified in the Pound-Rebka experiment are not gravitational at all!

\subsection{The ``Clock Hypothesis''}\label{clock}
The assumption, in both special and general relativity, that accelerating clocks are affected only by the time dilation associated with their instantaneous speed is widely referred to as the \emph{clock hypothesis}. It guarantees, in both theories, that the proper time registered by an ideal clock between events on its worldline, even when non-geodesic, is proportional to the length the relevant segment of the worldline obtained by integrating the spacetime metric interval along it. (By an ideal clock is meant a mechanical process which marches in step with the temporal parameter appearing in the fundamental equations of the non-gravitational interactions, when the mechanism is moving inertially.)
To call this a ``hypothesis'' is rather misleading. 

As Eddington nicely said of an accelerating clock:~``We may force it into its track by continually hitting it, but that may not be good for its time-keeping qualities.''\footnote{\cite[p.~64]{Edd2}.} Consider the Fe$^{57}$ nuclei in the Pound-Rebka experiment; although time dilation is negligible as we have seen, there is the question of what the effect of acceleration is on the radioactive nuclei. It is not just the acceleration $g$ of the sample that is relevant; thermal motion of the nuclei inside the sample involve accelerations of up to 10$^{16}g$! A back-of-the-envelope calculation made by Sherwin in 1960 showed that the mechanical effect of such lattice vibrations in distorting the nuclear structure is too small to appreciably affect the resonance M\"{o}ssbauer frequency.\footnote{\cite{Sherwin}.} A better-known test of the ``clock hypothesis'' was the famous g-2 muon experiment at CERN demonstrating relativistic time dilation in 1977, which involved accelerations of 10$^{18}g$.\footnote{\cite{muon}.} In an important paper published a decade later, Eisele used perturbation techniques in the theory of the weak interaction to calculate the approximate life time of the muons, and concluded that the correction to the clock hypothesis for such accelerations would be many orders of magnitude less than the accuracy of the 1977 experiment.\footnote{\cite{Eisele}.} (Eisele also noted that near radio-pulsars, magnetic fields are expected to exist of such strength as to lead to an acceleration-induced correction to muon decay of almost 1$\%$.) The calculations made by Sherwin and Eisele remind us that for a given acceleration, whether a particular clock ticks in accordance with the spacetime metric is not a matter of stipulation or luck, but depends crucially on the constitution of the clock. (The same goes for the influence of ``gravitational'' tidal effects within the clock, though this is not a practical concern within the solar system.) As Misner, Thorne and Wheeler emphasize:~``Velocity produces a universal time dilation; acceleration does not.''\footnote{See \cite[pp.~395-396]{MTW}; see also in this connection \cite{Giulini}, Remark 2.1. Note that this important insight is lost from view when an ideal clock in GR is \textit{defined} as one satisfying the clock hypothesis, rather than on the basis of its performance when moving inertially (freely falling).}

For any given clock, no matter how ideal its performance when inertial, there will in principle be an acceleration-producing external force, or even tidal effects inside the clock, such that the clock ``breaks'', in the sense of violating the clock hypothesis.\footnote{Similar considerations hold for the ``proper distance'' read off by accelerating rigid rulers. For further discussion of the role of rods and clocks in GR, see \cite{Brown2}. } Might it not be more appropriate to call it the \textit{clock condition}?

\section{The Einstein Equivalence Principle}\label{s3}

The operational significance of the metric field in GR is grounded in the following claims:~worldlines of massive test bodies are time-like geodesics; photons propagate along null geodesics; there is a link, as we have seen, between the proper time registered by an ideal clock and the metric (``chronometry'').\footnote{The metric is of course also surveyed by the proper distance read off by rigid rulers.} \textit{None of these claims is a direct consequence of Einstein's field equations.} The closest is the first of these, but geodesic motion only follows from the field equations when a natural condition is imposed on the stress-energy tensor associated with the test body (the so-called \emph{strengthened dominant energy condition}).\footnote{See e.g.~\cite{Malament}.}

The third claim (and maybe the second\footnote{Special relativity implies the existence of a finite invariant speed $c$; whether it is photons which instantiate this speed is a separate question; recently circumstances in which the group velocity of light \textit{in vacuo} is slightly less than $c$ were demonstrated experimentally. See \cite{Gio}.}) depends on special relativity being locally valid in GR. Soon after discovering his field equations, Einstein was aware that this is an hypothesis independent of both the field equations and the assumption that the metric has Lorentzian signature.\footnote{\cite{E3}.} Indeed, the local validity of special relativity is one component of what in the literature is often called the \textit{Einstein equivalence principle} (SEP).\footnote{Nomenclature varies here considerably; for example, SEP is related to the ``medium strong equivalence principle'' in \cite[\S3.2.4]{Ciuf}. In the philosophical literature, it is frequently referred to as the Strong Equivalence Principle.}
Here is how Misner, Thorne and Wheeler express the principle:\begin{quote}
... in any and every local Lorentz frame, anywhere and anytime in the universe, all the (nongravitational) laws of physics must take on their familiar special relativistic forms.\footnote{\cite[p.~386]{MTW}.}
\end{quote}
Local Lorentz coordinates associated with an arbitrary point (event) $p$ in curved spacetime are such that at $p$, $g_{\mu\nu,\rho}\left(p\right)=0$ and $g_{\mu\nu}\left(p\right) = \eta_{\mu\nu}$ in its diagonal form. In such a coordinate system,  geodesic motion at $p$ is inertial. (This does not make the coordinate system a freely falling frame, of course; these are described by a Fermi normal coordinate systems in which the above conditions hold along selected time-like geodesics.\footnote{See, e.g. \cite[pp.~182-183]{Hartle}; \cite[pp.~230-242]{Will}.}) For points in the neighbourhood of $p$, these conditions will no longer hold, and the deviation will depend on the Riemann curvature tensor which itself depends on second derivatives of the metric. However, in a ``local'' neighbourhood of $p$, where tidal effects are small enough -- and this depends on the nature of the experiment being performed and hence the details of the apparatus -- the spacetime can be approximated as flat.\footnote{For further discussion, see \cite[p.~352]{Knox}.} (The spacetime region involved in the Pound-Rebka experiment is just such an example. However, one can imagine other experiments being performed in the same region which \textit{are} sensitive to tidal effects\footnote{See \cite[\S16.5]{MTW}.} and relative to these experiments the region is no longer ``local'' in the usual sense.) But what does it mean that the non-gravitational laws of physics take on their special relativistic form in such a region? Here is how Hans Ohanian put the matter:
\begin{quote}
At each point of space-time it is possible to find a coordinate transformation such that the gravitational field variables can be eliminated from the field equations of matter.\footnote{\cite{Ohanian2}. Similar formulations in the recent literature can be found in e.g.~\cite[p.~169]{Brown} and \cite[p.~352]{Knox}.}

\end{quote}
This is very much in the spirit of the slogan that in GR, locally the gravitational field can be ``transformed away". The idea is that in the equations for the non-gravitational interactions at $p$, neither the curvature tensor nor its contractions appear, and that these equations take their simplest form in the local Lorentz coordinates.\footnote{This latter assumption itself is remarkable; it is saying that all the non-gravitational forces are Lorentz-covariant with respect to the same family of coordinates; the non-triviality of such universality was emphasised by Anderson \cite[\S10.2]{Anderson}.} (This way of putting things suggests that ``gravity'' is associated with curvature of spacetime, a notion we have met before and one not unheard of in GR textbooks!\footnote{See sections \eqref{equivalence} and \eqref{3.2} above, and e.g.~\cite[p.~218]{MTW}.} But it is a quite distinct notion from that of Einstein as described in section \eqref{Einstein} above. There is no definitive right or wrong here!) Note that the curvature-elimination condition is stronger than local Lorentz covariance. As it turns out, however, this view encapsulates what we call the \emph{third possible misconception of GR}:

\begin{description}
\item[Misconception 3:] In GR, the dynamical equations for non-gravitational interactions, in local Lorentz coordinates at a point $p$, {always} take a form in which neither the curvature tensor nor its contractions appear.
\end{description}

In fact, the situation here is subtle, as was noticed by Eddington as early as 1923.\footnote{\cite[p.~176]{Edd}. See also in this connection \cite{Hehl, Hehl2}.}
To illustrate, consider Maxwell's equations in Minkowski spacetime:
\begin{align}
F\indices{^{\mu\nu}_{,\nu}} &= J^\mu \label{max} \\
F_{\left[\mu\nu,\lambda\right]} &= 0 \label{max2}.
\end{align}
By manipulating \eqref{max} and \eqref{max2}, we obtain the second-order \emph{wave equation} for $F_{\mu\nu}$ in Minkowski spacetime:
\begin{equation}\label{wave}
F\indices{_{\mu\nu,\lambda}^{\lambda}} = 2 J_{\left[\mu,\nu\right]}.
\end{equation}
Typically, dynamical equations of GR are obtained from such equations by way of the so-called \emph{minimal coupling} procedure, according to which all partial derivatives in expressions valid in flat Minkowski spacetime are replaced with covariant derivatives, and the Minkowski metric is replaced by the metric $g_{\mu\nu}$ of Lorentzian signature which is a solution of Einstein's field equations:
\begin{equation}
R_{\mu\nu} - \frac{1}{2}R g_{\mu\nu} = 8\pi G T_{\mu\nu},
\end{equation}
where $R_{\mu\nu}$ and $R$ are the Ricci curvature tensor and scalar, respectively;  $G$ is the gravitational coupling constant and $T_{\mu\nu}$ is the stress-energy tensor for matter fields (which is taken to absorb the contribution of the cosmological constant). Thus, the Einstein-Maxwell equations in a generically curved spacetime of GR are written as:
\begin{align}
F\indices{^{\mu\nu}_{;\nu}} &= J^\mu \label{max3} \\
F_{\left[\mu\nu;\lambda\right]} &= 0 \label{max4}.
\end{align}
However, unlike the case of \eqref{max} and \eqref{max2}, the covariant derivatives in \eqref{max3} and \eqref{max4} do not commute, leading to a more complex wave equation for $F_{\mu\nu}$ in curved spacetime:
\begin{equation}\label{covwave}
F\indices{_{\mu\nu;\lambda}^{\lambda}} = 2 \left( F\indices{^\lambda_{[\nu}}R\indices{_{\mu]\lambda}} - R_{\mu\nu\sigma\lambda}F^{\sigma\lambda} + J_{\left[\mu;\nu\right]} \right).
\end{equation}

Second-order equations such as \eqref{covwave} pick up terms that depend on curvature,\footnote{Another example is the second-order wave equation for the 4-vector potential in vacuum electrodynamics, which picks up a term linear in the Ricci tensor; see \cite[p.~605]{TW} and \cite[pp.~388, 389]{MTW}.} \emph{and these terms cannot in general be made to vanish in local Lorentz coordinates at $p$}, in violation of the supposition that the matter field equations all take their ``special relativistic forms'' in such coordinates. (It should be noted that despite their formulation of the SEP above, Misner, Thorne and Wheeler emphasise this point, as do Poisson and Will more recently.\footnote{\cite[p.~390]{MTW}, and \cite[p.~248]{Will}.})

Importantly, this state of affairs does not threaten the postulated Lorentz covariance of these equations at $p$ expressed in Lorentz coordinates defined relative to $p$ -- and arguably this is all that the local validity of special relativity amounts to in GR.\footnote{For further discussion see \cite{us}.} In particular, such covariance guarantees that clocks built out of such matter survey the metric of spacetime locally, subject to their satisfying the ``clock hypothesis'' for the accelerations in question, and their mechanism being to a very good approximation insensitive to tidal effects. But it would be wrong to conflate the further principle of minimal coupling with the ability in all cases ``to eliminate gravitational field variables'' from the non-gravitational field equations by restriction to Lorentz coordinates, even when the curvature dependent terms may be insignificant in the experiments of interest. Accordingly, we may correct the third possible misconception in the following way:

\begin{description}
\item[Correction 3:] In GR, second- and higher-order dynamical equations for non-gravitational interactions, in local Lorentz coordinates at a point $p$ and constructed using the minimal coupling scheme, take a form in which the curvature tensor and/or its contractions appear.
\end{description}

\section{Conclusions and Outlook}

In this paper, we have elaborated what we take to be three possible foundational misconceptions in GR, which appear to have some instantiations in the literature. Given that such misconceptions go to the heart of concepts such as gravity, inertia, spacetime curvature, and inter-theory connections with both Newtonian mechanics and special relativity, perspicuity in these matters is of some importance. For this reason, 
we have in each case provided what we take to be a clear and unambiguous correction. The hope is that by doing so, a further modicum of conceptual clarity can be passed on to students of this most remarkable of theories, as we enter what Ferreira justly calls ``the century of GR''.\footnote{\cite{Pedro}.}

\section*{Acknowledgments}
We thank Dennis Lehmkuhl and Simon Saunders for very helpful discussions; Erik Curiel, Patrick D\"{u}rr, Ricardo Heras, Jos\'{e} W.~Maluf, Eric Poisson, Geoffrey Stedman, and David Wallace for comments on earlier drafts of this paper; and three anonymous referees for constructive comments which led to significant improvements. J.R.~is supported by an AHRC studentship and is also grateful to Merton College, Oxford for support.


\bibliographystyle{plain}

\end{document}